%%%%%
%%%
\input harvmac
\noblackbox

% -------------------------AK Definitions
\font\cmss=cmss10 \font\cmsss=cmss10 at 7pt
 \def\inbar{\,\vrule height1.5ex width.4pt depth0pt}
\def\IZ{\relax\ifmmode\mathchoice
{\hbox{\cmss Z\kern-.4em Z}}{\hbox{\cmss Z\kern-.4em Z}}
{\lower.9pt\hbox{\cmsss Z\kern-.4em Z}}
{\lower1.2pt\hbox{\cmsss Z\kern-.4em Z}}\else{\cmss Z\kern-.4em
Z}\fi}
\def\IB{\relax{\rm I\kern-.18em B}}
\def\IC{{\relax\hbox{$\inbar\kern-.3em{\rm C}$}}}
\def\ID{\relax{\rm I\kern-.18em D}}
\def\IE{\relax{\rm I\kern-.18em E}}
\def\IF{\relax{\rm I\kern-.18em F}}
\def\IG{\relax\hbox{$\inbar\kern-.3em{\rm G}$}}
\def\IGa{\relax\hbox{${\rm I}\kern-.18em\Gamma$}}
\def\IH{\relax{\rm I\kern-.18em H}}
\def\II{\relax{\rm I\kern-.18em I}}
\def\IK{\relax{\rm I\kern-.18em K}}
\def\IC{\relax{\rm I\kern-.18em C}}
\def\IR{\relax{\rm I\kern-.18em R}}

%----------------------------

\lref\gibbons{G.~W.~Gibbons,
``Vacuum Polarization And The Spontaneous Loss Of Charge By Black Holes,''
Commun.\ Math.\ Phys.\  {\bf 44}, 245 (1975).}
\lref\RS{L. Randall and R. Sundrum, ``An Alternative to Compactification''
Phys.\ Rev.\ Lett.\  {\bf 83}, 4690 (1999),
hep-th/9906064
}
\lref\itzhaki{N. Itzhaki, J. Maldacena, J. Sonnenschein, and
Shimon Yankielowicz, ``Supergravity and the Large N Limit of
Theories with Sixteen Supercharges'', hep-th/9802042.}
\lref\kut{D. Kutasov,``Orbifolds and Solitons,''
Phys.\ Lett.\  {\bf B383}, 48 (1996),
hep-th/9512145. }
\lref\sen{A. Sen,``Duality and Orbifolds,''
Nucl.\ Phys.\  {\bf B474}, 361 (1996), hep-th/9604070. }
\lref\smeared{See for example P.~Kraus, F.~Larsen and S.~P.~Trivedi,
``The Coulomb branch of gauge theory from rotating branes,''
JHEP {\bf 9903}, 003 (1999),
hep-th/9811120;
M.~Cvetic, S.~S.~Gubser, H.~Lu and C.~N.~Pope,
``Symmetric
potentials of gauged supergravities in diverse
dimensions and  Coulomb branch of gauge theories,''
hep-th/9909121.
}
\lref\stringglob{T. Banks, L. Dixon,
``Constraints On String Vacua With Space-Time Supersymmetry,''
Nucl.\ Phys.\  {\bf B307}, 93 (1988).
}
\lref\conifold{A. Strominger,
``Massless black holes and conifolds in string theory,''
Nucl.\ Phys.\  {\bf B451}, 96 (1995),
hep-th/9504090.
}
\lref\wittbound{E. Witten, ``Bound States of Strings and p-branes''
Nucl.\ Phys.\  {\bf B460}, 335 (1996),
hep-th/9510135.
}
\lref\singsolutions{
O.~DeWolfe, D.~Z.~Freedman, S.~S.~Gubser and A.~Karch,
``Modeling the fifth dimension with scalars and gravity,''
hep-th/9909134.
}
\lref\oferetal{O. Aharony, M. Berkooz, D. Kutasov, and
N. Seiberg, ``Linear Dilatons, NS5-branes, and Holography'',
hep-th/9808149.}
\lref\phases{E. Witten, ``Phases of N=2 Theories in Two Dimensions''
Nucl.\ Phys.\  {\bf B403}, 159 (1993),
hep-th/9301042.
}
\lref\hermann{H. Verlinde,
``Holography and compactification,''
hep-th/9906182.
}
\lref\eddiverse{E. Witten,
``String theory dynamics in various dimensions,''
Nucl.\ Phys.\  {\bf B443}, 85 (1995),
hep-th/9503124.
}
\lref\susswitt{L. Susskind and E. Witten,
``The holographic bound in anti-de Sitter space,''
hep-th/9805114.
}
\lref\gaugebulk{
A.~Pomarol,
``Gauge bosons in a five-dimensional theory with localized gravity,''
hep-ph/9911294;
;
H.~Davoudiasl, J.~L.~Hewett and T.~G.~Rizzo,
``Bulk gauge fields in the Randall-Sundrum model,''
Phys.\ Lett.\  {\bf B473}, 43 (2000),
hep-ph/9911262.
}
\lref\RSII{L. Randall and R. Sundrum,
``A large mass hierarchy from a small extra dimension,''
Phys.\ Rev.\ Lett.\  {\bf 83} (1999) 3370
hep-ph/9905221.
}
\lref\minsei{S. Minwalla and N. Seiberg,
``Comments on the IIA NS5-brane,''
JHEP {\bf 9906}, 007 (1999),
hep-th/9904142.
}
\lref\RSBH{
A.~Chamblin, S.~W.~Hawking and H.~S.~Reall,
``Brane-world black holes,''
Phys.\ Rev.\  {\bf D61}, 065007 (2000),
hep-th/9909205;
R.~Emparan, G.~T.~Horowitz and R.~C.~Myers,
``Exact description of black holes on branes,''
JHEP {\bf 0001}, 007 (2000),
hep-th/9911043;
S.~B.~Giddings, E.~Katz and L.~Randall,
``Linearized gravity in brane backgrounds,''
JHEP {\bf 0003}, 023 (2000),
hep-th/0002091.
}
\lref\tomofer{O. Aharony and T. Banks,
``Note on the quantum mechanics of M theory,''
JHEP {\bf 9903}, 016 (1999),
hep-th/9812237.
}
\lref\glob{M.~Kamionkowski and J.~March-Russell,
``Are textures natural?,''
Phys.\ Rev.\ Lett.\  {\bf 69}, 1485 (1992),
hep-th/9201063;
R.~Holman, S.~D.~Hsu, E.~W.~Kolb, R.~Watkins and L.~M.~Widrow,
``Cosmological texture is incompatible with Planck scale physics,''
Phys.\ Rev.\ Lett.\  {\bf 69}, 1489 (1992).}
\lref\adscft{J.~Maldacena,
``The large N limit of superconformal field theories and supergravity,''
Adv.\ Theor.\ Math.\ Phys.\  {\bf 2}, 231 (1998),
hep-th/9711200;
E.~Witten,
``Anti-de Sitter space and holography,''
Adv.\ Theor.\ Math.\ Phys.\  {\bf 2}, 253 (1998)
hep-th/9802150;
S.~S.~Gubser, I.~R.~Klebanov and A.~M.~Polyakov,
``Gauge theory correlators from non-critical string theory,''
Phys.\ Lett.\  {\bf B428}, 105 (1998),
hep-th/9802109.}
\lref\duff{M.~J.~Duff and J.~T.~Liu,
``On the equivalence of the Maldacena and Randall-Sundrum pictures,''
hep-th/0003237.}
\lref\gub{S.~S.~Gubser,
``AdS/CFT and gravity,''
hep-th/9912001.}
%n added two refs
\lref\addk{N.~Arkani-Hamed, S.~Dimopoulos, G.~Dvali and N.~Kaloper,
``Infinitely large new dimensions,''
Phys.\ Rev.\ Lett.\ {\bf 84}, 586 (2000),
hep-th/9907209.
}
\lref\rcm{R.~C.~Myers,
``Dielectric branes,''
JHEP {\bf 9912}, 022 (1999),
hep-th/9910053.
}
\lref\fivebrane{A.~Strominger,
``Heterotic Solitons,''
Nucl.\ Phys.\  {\bf B343}, 167 (1990);
C.~G.~Callan, J.~A.~Harvey and A.~Strominger,
``Supersymmetric string solitons,''
hep-th/9112030.}
\lref\graviphoton{
H.~Lu and C.~N.~Pope,
``Branes on the brane,''
hep-th/0008050.
M.~Cvetic, H.~Lu and C.~N.~Pope,
``Brane-world Kaluza-Klein reductions and branes on the brane,''
hep-th/0009183. 
M.~J.~Duff, J.~T.~Liu and W.~A.~Sabra,
``Localization of supergravity on the brane,''
hep-th/0009212; 
}
\lref\pomaral{
A.~Pomarol,
``Grand unified theories without the desert,''
hep-ph/0005293.
}
\lref\andyhet{
A.~Strominger,
``Superstrings With Torsion,''
Nucl.\ Phys.\  {\bf B274}, 253 (1986).
}
\lref\HW{
E.~Witten,
``Strong Coupling Expansion Of Calabi-Yau Compactification,''
Nucl.\ Phys.\  {\bf B471}, 135 (1996)
hep-th/9602070. 
}
\lref\edphases{
E.~Witten,
``Phases of N = 2 theories in two dimensions,''
Nucl.\ Phys.\  {\bf B403}, 159 (1993)
hep-th/9301042.
}
\lref\globalsymm{N. Kaloper, E. Silverstein, and L. Susskind,
``Gauge symmetry and localized gravity in M theory,''
hep-th/0006192.
}
\lref\domainwall{S. Kachru, M. Schulz, E. Silverstein 
%``Self-tuning flat domain walls in 5d gravity and string theory,''
Phys.\ Rev.\  {\bf D62}, 045021 (2000)
hep-th/0001206;
N.~Arkani-Hamed, S.~Dimopoulos, N.~Kaloper and R.~Sundrum,
``A small cosmological constant from a large extra dimension,''
Phys.\ Lett.\  {\bf B480}, 193 (2000)
hep-th/0001197.
}
\lref\hlz{G. Horowitz, I. Low, and A. Zee,
``Self-tuning in an outgoing brane wave model,''
Phys.\ Rev.\  {\bf D62}, 086005 (2000)
hep-th/0004206.
}
\lref\wiletc{
J.~L.~Feng, J.~March-Russell, S.~Sethi and F.~Wilczek,
``Saltatory relaxation of the cosmological constant,''
hep-th/0005276.
}
\lref\bp{R. Bousso and J. Polchinski,
``Quantization of four-form fluxes and 
dynamical neutralization of the  cosmological constant,''
JHEP {\bf 0006}, 006 (2000)
hep-th/0004134.
}
\lref\bd{
T.~Banks, M.~Dine and L.~Motl,
``On anthropic solutions of the cosmological constant problem,''
hep-th/0007206.
}
\lref\mn{
J.~Maldacena and C.~Nunez,
``Supergravity description of field 
theories on curved manifolds and a no  go theorem,''
hep-th/0007018.
}
\lref\tom{T. Banks, 
%``Cosmological breaking of supersymmetry or little Lambda goes back to  the future II,''
hep-th/0007146.
}
\lref\giant{J. McGreevy, L. Susskind, N. Toumbas, 
``Invasion of the giant gravitons from anti-de Sitter space,''
JHEP {\bf 0006}, 008 (2000)
hep-th/0003075.
}
\lref\elast{E. Silverstein, ``Extended States from Warped
Compactifications of M Theory'',
hep-th/0009057.
}
\lref\withallan{A. Adams and E. Silverstein, work in progress.}
\lref\reldef{
L.~Girardello, M.~Petrini, M.~Porrati and A.~Zaffaroni,
``Novel local CFT and exact results on perturbations of N = 4 super  Yang-Mills from AdS dynamics,''
JHEP {\bf 9812}, 022 (1998)
hep-th/9810126; 
D.~Z.~Freedman, S.~S.~Gubser, K.~Pilch and N.~P.~Warner,
``Renormalization group 
flows from holography supersymmetry and a  c-theorem,''
hep-th/9904017;
J.~Polchinski and M.~J.~Strassler,
``The string dual of a confining four-dimensional gauge theory,''
hep-th/0003136.
}
\lref\conifold{P.~Candelas and X.~C.~de la Ossa,
``Comments On Conifolds,''
Nucl.\ Phys.\  {\bf B342}, 246 (1990).}
\lref\sw{N.~Seiberg and E.~Witten,
``Electric - magnetic duality, monopole condensation, and confinement in N=2 supersymmetric Yang-Mills theory,''
Nucl.\ Phys.\  {\bf B426}, 19 (1994)
hep-th/9407087.
}
\lref\eddiverse{
E.~Witten,
``String theory dynamics in various dimensions,''
Nucl.\ Phys.\  {\bf B443}, 85 (1995)
hep-th/9503124.
}
\lref\andy{A.~Strominger,
``Massless black holes and conifolds in string theory,''
Nucl.\ Phys.\  {\bf B451}, 96 (1995)
hep-th/9504090.
}
\lref\smallinst{
E.~Witten,
``Small Instantons in String Theory,''
Nucl.\ Phys.\  {\bf B460}, 541 (1996)
hep-th/9511030.
}
\lref\kss{
S.~Kachru, N.~Seiberg and E.~Silverstein,
``SUSY Gauge Dynamics and Singularities of 4d N=1 String Vacua,''
Nucl.\ Phys.\  {\bf B480}, 170 (1996)
hep-th/9605036.
}
\lref\ks{
S.~Kachru and E.~Silverstein,
``Chirality-changing phase transitions in 4d string vacua,''
Nucl.\ Phys.\  {\bf B504}, 272 (1997)
hep-th/9704185.
}
\lref\topchange{
P.~S.~Aspinwall, B.~R.~Greene and D.~R.~Morrison,
``Calabi-Yau moduli space, mirror manifolds and spacetime topology  change in string theory,''
Nucl.\ Phys.\  {\bf B416}, 414 (1994)
hep-th/9309097;
J.~Distler and S.~Kachru,
``(0,2) Landau-Ginzburg theory,''
Nucl.\ Phys.\  {\bf B413}, 213 (1994)
hep-th/9309110;
B.~R.~Greene, D.~R.~Morrison and A.~Strominger,
``Black hole condensation and the unification of string vacua,''
Nucl.\ Phys.\  {\bf B451}, 109 (1995)
hep-th/9504145.
}
\lref\dealwis{
S.~P.~de Alwis,
``Brane world scenarios and the cosmological constant,''
hep-th/0002174.
}

\Title{\vbox{\baselineskip12pt\hbox{hep-th/0010144}
\hbox{SLAC-PUB-8671}
}}
{\vbox{\centerline{Gauge Fields, Scalars, Warped Geometry,}\smallskip
\centerline{and Strings}}
}

\centerline{ 
Eva Silverstein}
\bigskip
\bigskip
\centerline{Department of Physics ~~and  ~~ SLAC}
\centerline{Stanford University}
\centerline{Stanford, CA 94305/94309}
\bigskip
\medskip
\noindent

We review results on several interesting phenomena in warped
compactifications of M theory, as presented at
Strings 2000.  The behavior of gauge fields
in dimensional reduction from $d+1$ to $d$ dimensions in
various backgrounds is explained from the point
of view of the holographic duals (and a point raised
in the question session at the conference is addressed).  
We summarize the role of additional
fields (in particular scalar fields) in 5d warped geometries
in making it possible for Poincare-invariant domain wall solutions
to exist to a nontrivial order in a controlled approximation
scheme  
without fine-tuning of parameters in the $5d$ action
(and comment on the status of the singularities arising
in the general relativistic description of these solutions).  
Finally,
we discuss briefly the emergence of excitations of wrapped
branes in warped geometries whose effective thickness, as
measured along the Poincare slices in the geometry, grows
as the energy increases.

\Date{October 2000}

\newsec{Introduction}

Generic general-relativistic 
spacetime backgrounds with $d$-dimensional Poincare invariance
have a metric of the form

\eqn\genmet{
e^{2A(y)}\eta_{\mu\nu}dx^\mu dx^\nu+H_{IJ}(y)dy^I dy^J
}
where $\mu,\nu=0,\dots,d-1$.  Canonical examples in M theory 
with a nontrivial warp factor $e^{2A(y)}$ include
heterotic compactifications with $(0,2)$ worldsheet supersymmetry
\andyhet, 
compactified Horava-Witten theory \HW, $AdS_{d+1}$ and its
relevant deformations, linear dilaton theories such as for example 
the NS5-brane solution \fivebrane\ and the conifold singularity 
\conifold\edphases, 
and no doubt many more solutions yet to be discovered with less
supersymmetry.  

Many warped backgrounds have non-gravitational holographic
duals \adscft, but most have no known equivalent ``boundary theory''.  
Almost all of these backgrounds have curvature singularities
and/or strong coupling at some finite proper distance from a generic
points on the component of the geometry parameterized by
$y^I$, so that general relativity breaks down in this region
of the background.      

It is important and interesting to understand as precisely
as possible the physics of this type of background, in
particular to see if any new phenomena emerge from the
warped shape of the spacetime.  In this
talk I will review results on three aspects of this physics:

\noindent (1)  The behavior of gauge fields 

\noindent (2)  The role of for example scalar fields
in making possible, to the leading order in a controlled
approximation scheme,  solutions with Poincare invariance even after
some nontrivial quantum corrections to the vacuum energy have been
included, and 

\noindent (3)  The behavior of massive states coming from wrapped
branes in this sort of geometry:  in particular one finds a new
corner of the theory where excitations can be seen to grow
in size as they grow in energy as a consequence of the warping
in the metric \genmet.  

\newsec{Gauge Fields}

If we focus on cases where the warping occurs
along a single direction $y$, the low energy
effective action (to the extent that it is reliable) takes the form
\eqn\genac{S=\int d^dx dy\sqrt{g}\biggl(a(\phi)R+b(\phi)(\nabla\phi)^2
+c(\phi)F^2-\Lambda(\phi)\biggr).}

If the integral over $y$ of the Einstein term is finite one obtains
a finite $d$-dimensional Planck scale and ``trapped gravity'' \RS.
This holds also for a $d$-dimensional graviphoton that arises
from a $d+1$-dimensional two-form potential, as demonstrated recently in
\graviphoton.  On the other hand if there is a $d+1$-dimensional
vector potential, the dimensional reduction of its kinetic term might   
give a divergence independent of what is going on with the
graviton kinetic term. In particular in the $\int d^dxdy F^2$ term
there are two powers of the inverse $d+1$-dimensional metric
as opposed to the single power of $g^{\mu\nu}$ in the dimensional
reduction of the Einstein term.  

The coefficient of the $d$-dimensional
gauge kinetic term ${1\over e_d^2}\int d^dx tr F^2$  is the inverse
effective gauge coupling (charge) squared of the $d$-dimensional gauge
theory.  If the effective gauge coupling in $d$-dimensions
is zero, then one might naively infer that this symmetry behaves
like a global symmetry rather than a gauge symmetry after
dimensional reduction.  This would be very surprising since
the black hole no-hair theorems, at least in contexts where
they have been studied, indicate that information about global
charges is lost in processes in which the black hole absorbs
particles which carried this charge.  On the other hand in 
this context, where the symmetry is a bona fide gauge symmetry
in $d+1$ dimensions, the charges must be conserved.  

In \globalsymm\ we found
that in several known examples of warped geometries in which 
${1\over e_d^2}$ diverges, this divergence is either indicative
of a conventional screening effect or the effective theory
\genac\ breaks down at some $y<\infty$ where new behavior takes
over that also has a conventional behavior in the IR of
the $d$-dimensional description of the physics.     

The simplest example is the cutoff $AdS_5$ background studied
by Randall and Sundrum \RS.  In this background, $A(y)=-|y|/L$
where $L$ is the curvature radius of AdS.  The calculation giving
the effective charge in $4d$ is
\eqn\chargred{
{1\over e_4^2}\sim {1\over e_5^2}\int_0^{y_0}dy e^{-4y/L}(e^{2y/L})^2
\sim y_0}  
where we have introduced an IR momentum
cutoff $p_0$ through the relation $y_0\sim Llog(p_0L)$ following
from the metric.  The first factor in the integrand comes from
$\sqrt{g}$,  and the last from the two powers of the inverse
metric involved in forming $tr F^2$.  

So we are finding
\eqn\fincharge{
 {1\over e_4^2}\sim {1\over e_5^2} L log(p_0L)
}
This logarithmic IR divergence is the result one
would expect from screening of
the charge from the $4d$ description (as first pointed out to
us by E. Witten, and as first noted in papers of
Pomaral \pomaral).  This interpretation is confirmed by an explicit
calculation of the electrostatic potential arising from a point
source of charge at $y=0$:
\eqn\potcalc{
A_0(p, y=0)=-{{QK_1(pL)}\over{2 K_0(pL)}}
\rightarrow _{p\to 0} {Q\over{p^2 log p}}
}

Similarly one finds a generalization of the screening effect
to higher dimensions and higher-form fields.  In dimension $d$,
for a $q$-form field strength, as a function of IR momentum
cutoff $p_0$, we find  
\eqn\qcharge{\eqalign{{1\over e_q^2}\propto ~~~~~~
&e^{{R\over L}(2q-d)}\sim {1\over{p_0^{2q-d}}}~~~~2q\ne d\cr
&R\sim log(p_0)~~~~2q=d\cr}}

At the conference, M. Duff asked about the consistency of
this result with the possibility of dualizing $q$-form
field strengths to $5-q$-form field strengths.  For
example a scalar field $\eta$ with a 1-form field strength would
be dual to a 3-form potential field $C$ with a 4-form field
strength.  The latter, from \qcharge, gives a mode with
zero charge upon dimensional reduction; whereas a scalar
field, like gravity, is left with nontrivial interactions
after dimensional reduction.  

I believe the answer to this is as follows (this
result was obtained in collaboration with M. Schulz).  
The equation for dualizing a form, for example
$d\eta = *d C $, is a linear differential equation 
which locally has a solution.  There is no guarantee,
however, that this solution is nonsingular everywhere.

Consider the equation for a scalar field in the background
\genmet.  A massless mode in $4d$ satisfies the equation
\eqn\scaleq{
\eta''+4\eta' A'=0
}
where primes denote differentiation with respect to $y$.  
One obvious solution is the zero mode, $\eta=\eta(x)$
independent of $y$.  This is the mode which gets ``trapped'' upon
dimensional reduction, with a finite kinetic term.  As noted
by Duff in his question, this cannot be the solution dual to the
three-form potential.

There is another solution to \scaleq, which becomes
singular at the AdS horizon in the RS geometry.  Integrating
\scaleq, this solution satisfies
\eqn\nextscal{
\eta'=\hat\eta(x) e^{-4A(y)}
}
For the RS geometry, this yields
\eqn\scalfin{
\eta\sim \hat\eta(x)e^{+4y/L}
}  

The dimensional reduction of the kinetic term for this scalar
goes like
\eqn\scaldim{
\int_0^{y_0}dy e^{-4y/L}(e^{4y/L})^2\sim e^{4y_0/L}
}
from one power of $\sqrt{g}$ and two powers of $e^{+4y/L}$
from the solution \scalfin\ appearing in the quadratic action
for $\eta$.  

This is the same divergence which arises for the dimensional
reduction of the standard zero-mode solution for the three-form
potential $C$, which goes like
\eqn\threeform{
\int_0^{y_0} dy e^{-4y/L} (e^{2y/L})^4
}
from one power of $\sqrt{g}$ and four powers of the inverse
metric required to form the square of the four-form field
strength.  So the second solution \scalfin\ for the scalar
is evidently the one dual to the zero-mode of the three-form
potential.  

Consider the $N$-NS 5-brane solution of type II string 
theory \fivebrane.  It
has a string-frame metric and dilaton
\eqn\lindilone{\eqalign{
&ds^2=dx_6^2+dr^2+l_s^2Nd\Omega_3^2\cr
&\phi=\alpha r\cr
}}
with $\alpha=1/l_s\sqrt{N}$.

The string-frame ten-dimensional action is
\eqn\strac{\int d^6x dr d\Omega_3 \biggl[
e^{-2\phi}(R+(\partial\phi)^2)+K_{RR}^2\biggr]}
where $K$ is the field strength for the RR U(1) gauge field of
type IIA string theory, or the field strength for the 2-form
RR gauge potential of type IIB string theory.

In the dimensional reduction, $6d$ gravity survives (the $6d$
Planck scale ending up finite because of the coupling
of the Einstein term to $e^{-2\phi}$ in string frame).  
On the other hand the RR gauge field effectively propagates
in seven flat dimensions according to the metric
\lindilone\ since its kinetic term is
independent of $\phi$.  This is not a screening phenomenon
in $6d$.  

The resolution is that the breakdown of the solution
\lindilone\ down the throat due to strong coupling effects
is important.  The IIA NS5-brane is fundamentally an M5-brane
at a point on the eleventh circle of M 
theory \itzhaki\minsei.  This means
that deep in the IR region of the solution the RR gauge
symmetry is spontaneously broken, and this Higgs mechanism
wards off the more exotic possibility of a conserved
global symmetry persisting in $6d$.  

These examples provide further evidence for the robustness of
the arguments against global symmetries in quantum gravity.
However, it is interesting to keep one's eyes out for more
exotic examples that might arise in which a naive divergence
in the dimensionally reduced gauge kinetic term persists into
the infrared in a way that cannot be understood from 
screening or the Higgs mechanism.
In such a background, the black-hole no-hair theorems would
need to be analyzed carefully.

\newsec{Scalars}

Let us focus now on the physics of gravity plus scalars, with action
in a $5d$ bulk and on a $4d$ $\delta$-function localized brane
given by

\eqn\basicac{\eqalign{
S=
&\int d^5x\sqrt{-G}\biggl[
R-{4\over{3}}(\nabla\phi)^2-\Lambda e^{a\phi}\biggr]\cr
&+\int d^4x\sqrt{-g}(-f(\phi))\cr
}
} 
We have taken the bulk $\Lambda$ to be zero to a leading approximation,
having in mind bulk supersymmetry which is only broken at the level
of interactions with the brane.  

In contrast to the Randall-Sundrum system, this one has $4d$ 
Poincare-invariant solutions for {\it generic} brane tension $f(\phi)$
\domainwall.
The solutions in the bulk are:
\eqn\scal{
\phi(y)=\pm{3\over 4} log|{4\over 3}y+c|+d
}
\eqn\warp{
A(y)={1\over 4} log|{4\over 3}y+c|+\tilde d 
}
At $y=-{3\over 4}c$, this configuration has curvature
singularities:  general relativity breaks down near these
points.  

Einstein's equations at the wall at $y=0$ (which boil down
to Israel matching conditions there) are solved by adjusting
the values of integration constants $c,d$, {\it not} by tuning
parameters like $f(\phi)$ in the Lagrangian.  So quantum
corrections to $f(\phi)$ will not ruin the fact that
there is a flat solution.  
This is perhaps encouraging, since in M theory all indications
are that one does not
have the freedom to tune parameters in the theory (there
being no arbitrary dimensionless couplings put in);
we at most have the freedom to choose among
different solutions of the basic equations (still to be
determined!) of the theory.  

It is therefore an interesting goal
to exhibit as a first step a background of M theory with
near-vanishing cosmological term after nontrivial quantum
corrections are included, as also noted recently
in \dealwis.  The next (and presumably most difficult)
step is to understand why we live in such a background as 
opposed to one of the millions of others with large spacetime
curvatures or otherwise unrealistic low-energy
physics.

In our examples \basicac\scal\warp, we must first understand
the physics of the singularities that appear in the general-relativistic
description of the system.  Progress was made on this front
in two directions.  Firstly, Horowitz, Low, and Zee found stringy
cosmological solutions with again effectively zero cosmological
term independent of the parameters in the Lagrangian \hlz.  In these
solutions, there are again curvature singularities but they
are null, so
that there is no issue of additional boundary conditions 
needed for modes emanating outward from the brane.

Secondly, the static solutions above are in some
ways analogous to the types of backgrounds that have been
intensely studied recently as deformations of the AdS/CFT correspondence
to confining theories (see for example \reldef).  
In those backgrounds, a singularity in
the general relativistic approximation corresponds to the presence
of a mass gap beyond which there are no excitations of the field theory.
Just as the presence of a mass gap is generic to asymptotically 
free quantum field
theory, the presence of a singularity in the $5d$ gravity dual
in the (bad) general relativistic approximation is generic.
That is not to say that all singular backgrounds of this form
are resolved by quantum gravity effects.  Indeed, in this AdS/CFT context
the field theory side can only vouch for a discrete subset
of the continuum of apparent singular solutions in the GR description,
since at finite $N$ the field theory has a finite number of
vacua.
Luckily there are of order $e^{\sqrt{N}}$ such vacua.   

So the upshot, in this context, is this.  Couple
a large-N gauge theory to the standard model in the way
that is determined by the addition of a thin domain wall
to cut off the UV end of the gravity dual to the gauge 
theory.  Generically, one can choose a vacuum of the
large-N theory to zero the $4d$ cosmological term in the full
system, up to corrections that are parametrically suppressed
relative to the $TeV^4$ contribution expected from
standard model loops.  These corrections could also
introduce instabilities in the system, which is a
problem endemic to these models as well as those of \RS
(not to mention any string compactification with approximate
moduli).    

Is there any reason this vacuum is
preferred?  Aside from the partially anthropic arguments 
presented in \bp\wiletc (where another interesting procedure for finding
solutions of M theory with near-vanishing cosmological
term was obtained) I know of no argument for this at present.
One advantage of our setup in the context of the 
Brown-Teitelboim style analysis of \bp\ is that the
``discretuum'' of different vacua in our case all have
manifestly similar ``standard model'' physics since
this resides on the brane and the bulk is where the
cancellation mechanism arises.  As emphasized in 
\bd, this is an important consideration in entertaining
this sort of anthropic explanation for the history.  

However, it is worth emphasizing that the actual solutions
\scal\warp\ considered in \domainwall\ do not asymptote
to AdS space in bulk, but instead tend toward
flat space far from the singularity.  Therefore they are not dual
to a quantum field theory in toto.  It is possible that the
regime near the singularity does have an effective quantum
field theory dual, since the warp factor decreases there \mn.
In any case it is very interesting to try to understand
the physics of warped geometries which do not have a quantum
field theory dual.  Related to this is the question of whether
any backgrounds of this type can have sensible physics
in which effective quantum field theory breaks down so
that long-distance quantities like the cosmological constant
might be affected naturally by high-energy excitations in
the theory.

I cannot resist adding here more general comments on 
naked singularities in string theory.   
At such singularities, general relativity breaks down.
To me this is one of the most interesting features in
a spacetime, since it is an opportunity to
learn about physics of M theory that goes beyond
the long-wavelength general relativistic approximation.  
The resolution of singularities on
the Coulomb branch of gauge theories in Seiberg-Witten
theory \sw, in ADE limits of compactifications of 
type II string theory on K3 \eddiverse, at type II conifold
points \andy\ and their heterotic cousins \smallinst\kss\ks
(just to name a few examples),
involved in a detailed way the physics of
non-perturbative excitations of the theory and nontrivial
information about its strong-coupling behavior.  Without
these many resolutions of general-relativistic naked singularities,
the web of dualities relating different limits of M theory
would not exist.  One of the immediate applications of
such resolutions has also been the analysis
of controlled topology-changing transitions \edphases\topchange\ks\
in M theory.  

For the most general solutions described
in this section, it remains to be seen whether or not the singularities
have a (perhaps discretized) resolution, and if so whether a standard
$4d$ effective field theory arises at long distances.  
(In the above examples, on the order of a decade passed
between the original nakedly singular solutions being
written down (for the conifold for
example in \conifold) and the eventual quantum resolution; it
is not clear in our case when all the necessary ingredients
will be available to answer this question.) This is an interesting
open question.  

\newsec{Wrapped Branes}

One of the features of quantum gravity emphasized recently
for example in \giant\tom\ is the fact that at the highest energies,
excitations (namely black holes) grow in size as they grow
in energy (mass).  This is in stark contrast to the 
size=1/momentum uncertainty relation of elementary excitations
in quantum field theory.  This makes it conceivable that
such high-energy excitations could naturally 
affect long-distance parameters.  

It turns out that warped compactifications constitute another
context in which some excitations grow in size as they grow
in energy.  In general given a metric \genmet, there will
be wrapped branes on the compact component of the geometry
whose mass $m_0(y)$  will depend on $y$ since the volume of the cycle
on which the brane is wrapped varies as a function of $y$.
As measured along the Poincare slices, the energy will be
(taking into account the warp factor as well as the 
variation of the cycle volume)
\eqn\genmass{
E(y)\sim \sqrt{g_{00}(y)}m_0(y) 
}
On the other hand, because of the warp factor, the thickness
of the object is rescaled from its proper thickness $r_0$ because
of the warp factor:
\eqn\gensize{
R\equiv
\delta x_{||} \sim {1\over\sqrt{g_{ii}}}r_0
}

In \elast\ concrete examples were exhibited in which both $R$ and $E$ grow
in the same direction, so that
\eqn\reln{
E=TR^q
}
for some power $q>0$.  The spectrum of excitations of these
wrapped branes was worked out in a limit where a Kaluza-Klein
analysis was valid.  

Any situation in which there is this sort of growth of size with energy
suggests that interesting nonlocal effects might emerge from the dynamics
of these excitations.  This is under investigation \withallan.

\centerline{\bf Acknowledgements}
The work in \S2\ was done in collaboration with 
N. Kaloper
and L. Susskind.  The work reviewed in \S3\ was done in
collaboration with S. Kachru and M. Schulz, and 
concurrently by N. Arkani-Hamed, S. Dimopoulos, N. Kaloper, and R. Sundrum.
M. Duff asked an interesting 
question about the material in \S2.  
I thank all of the above, as well as those mentioned in the corresponding
papers, for many interesting discussions on these and related topics.
Finally I would like to thank the organizers of Strings 2000
for a stimulating conference.  
This work was supported by a
DOE OJI grant, by the A.P. Sloan
Foundation, and
by the DOE under contract DE-AC03-76SF00515.

\listrefs

\end